\documentclass[12pt]{article}

\usepackage{hyperref}
\usepackage{cite}
\usepackage{units}
\usepackage{epsfig}
\usepackage{pifont}
\usepackage{amsmath,amsfonts,amssymb,amsthm,mathrsfs}
\usepackage{graphicx,chngcntr,slashed}
\usepackage{cancel}
\usepackage{pstricks}
\usepackage{afterpage}
\usepackage{braket}
\usepackage{caption}
\usepackage{subcaption}
\usepackage{xcolor}
\usepackage{verbatim}
\usepackage{multirow}
\usepackage{tabularx,caption}
\usepackage{float}
\usepackage{afterpage}
\usepackage{comment}

\newcommand{\as}{\alpha_{\mathrm{s}}}

\newcommand{\LT}{\mathrm{T}}

\newcommand{\Lc}{\mathrm{c}}

\newcommand{\Lf}{\mathrm{f}}

\newcommand{\GeV}{\ \mathrm{GeV}}

\makeatletter
\def\section{\@startsection {section}{1}{\z@}{+3.0ex plus +1ex minus
  +.2ex}{2.3ex plus .2ex}{\large\bf\boldmath}}
\def\subsection{\@startsection{subsection}{2}{\z@}{+2.5ex plus +1ex
minus +.2ex}{1.5ex plus .2ex}{\normalsize\bf\boldmath}}
\def\subsubsection{\@startsection{subsubsection}{3}{\z@}{+3.25ex plus
 +1ex minus +.2ex}{1.5ex plus .2ex}{\normalsize\it}}

\graphicspath{{.}{plots/}}

\begin{document}
\thispagestyle{empty}

\def\thefootnote{\fnsymbol{footnote}}

\begin{flushright}
\end{flushright}


\newcommand\snowmass{\begin{center}\rule[-0.2in]{\hsize}{0.01in}\\\rule{\hsize}{0.01in}\\
\vskip 0.1in Submitted to the  Proceedings of the US Community Study\\ 
on the Future of Particle Physics (Snowmass 2021)\\ 
\rule{\hsize}{0.01in}\\\rule[+0.2in]{\hsize}{0.01in} \end{center}}

\snowmass

\vspace{1cm}

\begin{center}

{\Large {\bf Future prospects for parton showers}}
\\[3.5em]
{\large
Neda~Darvishi$^{1,2}$, 
Joshua Isaacson$^{3}$, 
M.R.~Masouminia$^{4}$, 
Zoltan~Nagy$^{5}$, 
Peter~Richardson$^{4}$, 
Davison~E.~Soper$^{6}$, 
}

\vspace*{1cm}

{\sl
$^1$ Department of Physics and Astronomy, Michigan State University, East Lansing, MI 48824, USA\\[1ex]
$^2$ Institute of Theoretical Physics, Chinese Academy of Sciences, Beijing 100190, China\\[1ex]
$^3$ Fermi National Accelerator Laboratory, P.~O.~Box 500, Batavia, IL 60510, USA\\[1ex]
$^4$ Institute for Particle Physics Phenomenology, Durham University, Durham DH1 3LE, United Kingdom\\[1ex]
$^5$ Deutsches Elektronen-Synchrotron DESY, 
 Notkestr.\ 85, 22607 Hamburg, Germany\\[1ex]
$^{6}$ Institute for Fundamental Science,
University of Oregon,
Eugene, OR  97403-5203, USA\\[1ex]
}

\end{center}

\vspace*{2.5cm}
\setcounter{page}{0}
\setcounter{footnote}{0}

\begin{abstract}
In this brief Snowmass White Paper for the Theory Frontier, we argue that there have been important recent developments in the algorithms used to generate a simulated parton shower and that further progress can be achieved in the coming decade. A much more detailed exposition can be found in a corresponding White Paper of the Energy Frontier.
\end{abstract}
\newpage



\section{Introduction}
\label{sec:introduction}

Parton shower event generators have proven to be very important since their introduction in the 1980s \cite{EarlyPythia, Gottschalk, EarlyHerwig}. Discussion of these tools and of their importance for the planning and interpretation of experiments is included in a White Paper of the Energy Frontier. In this White Paper for the Theory Frontier, we outline important developments in the algorithms used to generate a simulated parton shower. 

These algorithms are based on a detailed understanding of the structure of the QCD. In recent years there has been substantial work to translate our knowledge of this structure into practical computer algorithms. This work is closely related to developments in extending perturbative calculations of important hard processes to higher perturbative order. The work is also closely related to work to rearrange QCD perturbation theory for processes that contain large logarithms in their perturbative expansion. 

\section{Mini-review of parton showers}
\label{sec:review}

Parton shower event generators for hadron collisions, for example \textsc{Herwig}  \cite{Herwig}, \textsc{Pythia} \cite{Pythia}, and \textsc{Sherpa} \cite{Sherpa}, perform calculations of cross sections according to an approximation to the standard model or its possible extensions.

The main ideas behind parton shower event generators were developed in the 1980s \cite{EarlyPythia, Gottschalk, EarlyHerwig}. There has been extensive development of the algorithms since then \cite{dipolesG, lambdaR, dipolesGP, Herwig1992, Pythia1994, SherpaAlpha, SjostrandSkands, NSI, NSRingberg, Geneva}, leading to  programs that are quite sophisticated \cite{Herwig, Pythia, Sherpa, DinsdaleTernickWeinzierl, SchumannKrauss, PlatzerGiesekeI, PlatzerGiesekeII, Deductor, HochePrestel, Vincia2016}. Reviews of the field can be found in \cite{review2011, SjostrandReview}.

One can view a parton shower algorithm as beginning with the theorem \cite{factorization} that allows us to write a cross section for an infrared safe observable as a convolution of a hard scattering factor with parton distribution functions. Then the parton shower fills in more detail by using the renormalization group, using the insight that the scattering process appears differently depending on the hardness scale at which one examines it. The parton shower develops with decreasing values of a scale parameter that measures the hardness of interactions.\footnote{\textsc{Herwig}, however rearranges the ordering of splittings in its default parton shower so that larger angle splittings come first.} At the hardest scale, the scale of the hard interaction, there are just a few partons. Then, as the hardness scale at which we examine the process decreases, these partons split, making more partons in a parton shower. Thus, with respect to initial state partons, the shower evolution starts from the hard interaction and moves backward in time to softer initial state interactions. With respect to final state partons, shower evolution moves forward in time. At any stage, a certain amount of structure has emerged, while softer structure remains unresolved.

The parton shower uses functions called splitting functions that represent the probability for partons to split at a scale $\mu^2$ that is smaller than the scales of splittings that have come before. Thus the splitting functions represent the infrared singularity structure of QCD. At the current frontier of algorithm development, the splitting functions are evaluated at first order in an expansion in the strong coupling $\as(\mu^2)$. Although we speak of cross sections and probabilities in explaining the parton shower, we must remember that QCD is a quantum theory, so that we need to properly include quantum interference. This is especially relevant with respect to color and spin, which are intrinsically quantum variables.

As the resolution scale $\mu^2$ becomes smaller than a scale $\mu_\Lf^2$ of order $1 \GeV^2$, $\as(\mu^2)$ is not small enough to justify the use of perturbation theory to describe the development of the shower. A parton shower event generator aims to generate complete events, so it is necessary to turn to a model of physics at scales smaller than $\mu_\Lf^2$. This model produces hadrons from the partons produced in the perturbative part of the shower.

In the following sections, we turn to some of the topics of recent and future development in parton shower event generators.

\section{Matching and merging}
\label{sec:matching}

A parton shower is initiated by a hard scattering process, say $g g \to \mathrm{Higgs}$. If the hard scattering is calculated at lowest order, then the result is conceptually simple: the parton shower provides an approximated version of higher order corrections. However, if we want to use a hard process calculated at, say, NLO, then we need to subtract the approximated NLO corrections generated by the shower. The current state of the art for matching is the matching of NNLO QCD fixed-order computations with parton showers (NNLOPS). Matching techniques beyond NLO necessarily require accounting for different jet multiplicities. For instance, a NNLOPS result for $gg\to $ Higgs production must be NNLO accurate for inclusive Higgs production, but also NLO accurate for $gg \to$ Higgs + jet, and LO accurate for $gg \to$ Higgs + 2 jets. There has been a large amount of work on matching in recent years and work is ongoing \cite{MCatNLO, PowHeg1, PowHeg2, MENLOOPS, PowHegBox, FxFx, Hamilton:2012np, Hamilton:2012rf, MEPSatNLO, UNLOPS, PlatzerMatching, Hamilton:2013fea, MG5aMCNLO, GENEVANNLO, Alioli:2021qbf, Alioli:2020qrd, Alioli:2019qzz,
Czakon, KrkNLO2015, KrkNLO2016, MINLO3, HocheMatching1, HocheMatching2, PrestelN3LO,
Karlberg:2014qua, Bizon:2019tfo, Astill:2016hpa, Astill:2018ivh, Re:2018vac, Monni:2019whf, Monni:2020nks,   Lombardi:2020wju, Lombardi:2021wug, Lombardi:2021rvg, Buonocore:2021fnj, Zanoli:2021iyp, Mazzitelli:2020jio, Mazzitelli:2021mmm}.

We may also want to consider together two different processes, each calculated at beyond lowest order (for instance $g g \to \mathrm{Higgs}$ and $g g \to \mathrm{Higgs} + \mathrm{jet}$). Then each hard process can initiate a parton shower. Evidently, there is a certain amount of ambiguity in exactly how these processes should be combined. Again, there has been a large amount of work on this in recent years and work is ongoing \cite{CKKW, CKKWL, MLM, MENLOOPS, FxFx, Hamilton:2013fea, MEPSatNLO, UNLOPS, PlatzerMatching, GENEVANNLO, Alioli:2021qbf, Alioli:2020qrd, Alioli:2019qzz, MINLO3, HocheMatching1, HocheMatching2}.



\section{Quantum interference and dipoles}
\label{sec:dipoles}

Parton shower algorithms need to respect quantum mechanics. That is, one should consider the evolution of the quantum amplitude. One of the earliest shower algorithms, the one in \textsc{Herwig}, was invented to do just this in an approximate way. Most modern parton showers are of the ``dipole'' sort, so that one includes (with approximations) emissions from both members of the dipole, including the interference between emission from one member of a dipole and emission from the other. In this way, it is the quantum amplitudes that evolve.

In the description of gluon emission in a dipole shower, emission from parton $l$ interferes with emission from emission from a dipole partner parton $k$. In one way of arranging this in an algorithm, usually denoted an {\em antenna shower}, the whole $l$-$k$ dipole is considered as a unit \cite{Vincia2016}. More commonly, in a {\em partitioned dipole shower}, the emission process is partitioned into two parts \cite{Pythia, Sherpa, Deductor, Dire, PanScales}. When the emitted gluon is closer to being collinear with parton $l$, it is that parton that is treated as being the emitter in the treatment of momentum sharing in the emission. An important challenge in current research is to improve the approximations used to implement quantum interference in a dipole shower, especially with respect to spin and color.

\section{Spin}
\label{sec:spin}

Partons carry quantum spin. The spin of gluons affects the dependence of their splitting functions on the azimuthal angle of their decays. Thus one should keep track of quantum spin in the parton amplitudes. The alternative is to average over spins at each step, thus losing information that affects azimuthal angle distributions. For certain technical reasons, the implementation of quantum spin in a parton shower algorithm is much easier than the implementation of color: for spin, one can use the Collins-Knowles algorithm \cite{Collins:1987cp, Knowles:1987cu, Knowles:1988hu, Knowles:1988vs} However, the needed implementation has been lacking in many parton shower programs. 

\textsc{Herwig~7} (since v7.2.0~\cite{Bellm:2019zci}) has adopted the Collins-Knowles algorithm in both its default and dipole QCD parton shower schemes to include the spin-correlations between the production and decay of heavy particles, based on the developments introduced in~\cite{RichardsonSpin, Richardson:2001df}. This algorithm allows for the inclusion of the spin-correlations while maintaining the step-by-step approach of the event generation process. Further improvement in the treatment of heavy-quark spin distributions, through the inclusion of heavy quark effective field theory in \textsc{Herwig~7}'s cluster hadronization model and its heavy hadronic decay modes, will be introduced with the \textsc{Herwig~7.3} public release.

Spin correlations can also be included in a dipole shower using the Collins-Knowles algorithm \cite{NSspin}. Recent work \cite{Karlberg:2021kwr, Hamilton:2021dyz} has studied this in some detail. 

\section{Color}
\label{sec:color}

Partons carry quantum SU(3) color. This means that there is a bra amplitude describing many partons with their color and a ket amplitude for many partons with their color. Gluon emissions change the color state, as do virtual gluon exchanges. This is not so easy to describe in a computer because the color space for, say, twenty partons has approximately $10^{36}$ dimensions. Furthermore, an approximate version of virtual diagrams is included in a parton shower as an exponential, the Sudakov exponential. One can, of course, exponentiate a matrix on a computer, but not in $10^{36}$ dimensions.

Most programs use what is called the leading color (LC) approximation, which gives the leading term in an expansion in powers of $1/N_\Lc^2$ (with $N_\Lc = 3$). There is an improved version called the LC+ approximation \cite{NScolor}. This is an approximation to the proper evolution of the color amplitudes, but is still a rather crude approximation. Work to do better is currently ongoing. One method expands perturbatively in the difference between the full color splitting functions and their LC+ approximation \cite{NSColorTheory}. There has been extensive recent work \cite{Platzer:2012np, Bellm:2018wwz, Platzer:2018pmd, Isaacson:2018zdi} that studies color in parton shower evolution beyond leading color, including work \cite{AngelesMartinez:2018cfz, Forshaw:2019ver, Forshaw:2020wrq, DeAngelis:2020rvq, Holguin:2020joq, Hoche:2020pxj, Platzer:2020lbr} that accounts approximately for color amplitude evolution resulting from both real emission graphs and virtual exchange graphs. Ref.~\cite{HamiltonShowerSum} has worked to improve the treatment of color in parton showers without tying the description to the color amplitudes.

\section{Summation of large logarithms}
\label{sec:largelogs}

Many cross sections that play a role in particle physics depend on two very different momentum scales. In consequence, the coefficient of $\as^n$ in the perturbative expansion of such a cross section will contain powers of the logarithm of $L$ of the ratio of these scales. Typically, we find contributions proportional to $\as^n L^{2n-1}$ or $\as^n L^{2n}$. An example is the cross section to produce a virtual photon with squared momentum $Q^2$ and with transverse momentum $k_\LT$, with $L = \log(Q^2/k_\LT^2)$. The large logarithms $L$ spoil the usefulness of fixed order perturbation theory in calculating the cross section. There has been a very substantial theoretical effort over the years to sum the perturbative contributions that contain the most powers of $L$. For instance, soft-collinear-effective theory (SCET) has often been used for this purpose in recent years. 

This sort of analytical large logarithm summation is adapted to a particular observable cross section. On the other hand, a parton shower event generator samples many simulated events and allows the user to measure {\em any} cross section involving the resulting partons (or hadrons if one applies a hadronization model). Thus a parton shower is much more flexible than a dedicated calculation of the same cross section. Furthermore, a parton shower uses parton splitting functions that contain the soft and collinear singularities of QCD, so it has the potential to sum the large logarithms correctly.

Does it? In some cases, it does \cite{NSdglap, NSZpT}. However, the answer depends on what the details of the parton shower are and what logarithms one would like to sum. Clearly, it is important to understand this connection better. There has been interesting recent work on this subject \cite{PanScales, HamiltonShowerSum, Forshaw:2020wrq, Nagy:2020rmk } and we can anticipate more results in the near future.

\section{Threshold logarithms}
\label{sec:thresholdlogs}

There is one class of large logarithms that is typically not included in parton shower algorithms. These are the ``threshold logarithms'' that occur in hard scattering cross sections at hadron colliders \cite{Sterman1987}. These logarithms can be thought of as arising from a mismatch between the kinematic limits in the DGLAP evolution equation for parton distribution functions and the kinematic limits of splittings in a parton shower or in the contributions to the theoretical cross section beyond leading order. The effects of threshold logarithms are often important, so there has been an extensive effort over the years to analyze them analytically. 

The effect of threshold logarithms can be incorporated in a parton shower algorithm \cite{NSThreshold, NSThresholdII}. Typically, this effect is left out of parton shower event generators, although the first perturbative term in the threshold log summation is included if the parton shower is matched to an NLO perturbative calculation of the hard scattering cross section. One can anticipate that a threshold factor will be included in more parton shower event generators so as to improve their accuracy in the future.

\section{NLO shower}
\label{sec:NLOshower}

Current parton shower algorithms are based on parton splitting probabilities calculated at lowest order in the strong coupling, order $\as^1$. One might hope \cite{NSAllOrder} to have a parton shower based on parton splitting probabilities calculated at order $\as^2$ and beyond. There has been recent progress in this direction \cite{JadachJHEP, JadachPolonica, HartgringNLO, SkandsNLO, HocheNLO1, HocheNLO2, HocheNLO3, PlatzerNLO}. The theoretical issues that must be addressed are similar to the issues involved in designing the subtractions for a perturbative calculation of jet cross sections at next-to-next-to-leading order. One can have the real emission of two partons or the emission of one parton with one virtual exchange or one can have two virtual exchanges. In the case of, say, two gluons, the two gluons can be nearly collinear with one existing parton or one gluon can be collinear with one parton and one collinear with another parton. Alternatively, one or both gluons can be soft, leading to interference between soft gluons from up to four different radiators. In order to achieve a consistent calculation, one must subtract the results from an order $\as^1$ parton shower prediction, expanded up to order $\as^2$, as well as remove the overlap between collinear and soft approximations to the matrix element. This can be achieved using phase space sectorization, partial fractioning, explicit subtractions or a combination thereof. The precise technology is often inspired by the aim to provide a parton shower that can be matched to a fixed-order NNLO calculation, using an extension of the MC@NLO or POWHEG technique \cite{PreussNNLO}.  For emissions from an initial state parton, the parton distribution functions and their evolution is involved. Although a complete algorithm lies in the future, we believe that the effort holds great promise.

\section{Electroweak radiation in a parton shower}
\label{sec:EWPS}

A consistent inclusion of electroweak (EW) radiation in a QCD dominant parton shower, or in the parton shower with incoming leptons, is a theoretical challenge. Issues arise because the SU(2) gauge theory is spontaneously broken, so that there is a transition in the behavior of the theory between low scales and very high scales. Furthermore, hadrons (and leptons) are singlets under the color SU(3) gauge group but carry charge with respect to the electroweak gauge group. This makes the standard collinear factorization that applies for QCD problematic for electroweak shower evolution. 

However, EW effects can be important at high energies, both for the LHC and for future colliders. In hadron-hadron collisions virtual EW radiative corrections from EW gauge boson self-interactions are usually large and destructive, and in a region of the phase-space where these become important, one has to consider real EW emissions of the EW bosons to counter these virtual effects. Also, one expects that the heavy EW bosons could be considered as massless partons and contribute to parton shower splittings when probing high-enough energy scales. In such a study, one is required to employ a process-independent EW parton shower scheme that is capable of performing interleaved QCD and EW radiations on equal footing. 

There has been substantial work to address these issues \cite{Chen:2016wkt,Bauer:2017isx, Bauer:2017bnh, Manohar:2018kfx, Bauer:2018xag, Baumgart:2018ntv, Billis:2019evv, Cuomo:2019siu, Kleiss:2020rcg, Han:2020uid, Huang:2020iya, Jung:2021mox, Han:2021kes, Buarque:2021dji, Gellersen:2021caw, Ruiz:2021tdt, Masouminia:2021kne,Darvishi:2021het, Chiesa:2013yma, Christiansen:2014kba, Krauss:2014yaa, Mangano:2002ea, Brooks:2021kji}.

Recently, such a practical approach has been implemented in \textsc{Herwig}'s default angularly-ordered parton shower algorithm, covering the full scope of final-state EW radiations in addition to the initial-state EW gauge and Higgs boson emissions~\cite{Masouminia:2021kne, Darvishi:2021het}. This {QCD$\oplus$QED$\oplus$EW} parton shower, however, does not include initial-state EW gauge boson self-interactions since this would require the development of reliable EW parton distributions functions. This parton shower algorithm has been subjected to numerous validation tests in both performance~\cite{Masouminia:2021kne} and physics~\cite{Darvishi:2021het}. Particularly, this algorithm has been utilized in the precision calculation of the possible signature of the Maximally Symmetric multi-Higgs-doublet model (MS-$n$HDM)~\cite{Darvishi:2019ltl,Darvishi:2020teg,Birch-Sykes:2020btk,Darvishi:2021txa,Darvishi:2022zag}, for the simplest case i.e. MS-2HDM~\cite{Darvishi:2020paz}. 

It has been shown that the {QCD$\oplus$QED$\oplus$EW} parton shower scheme improves \textsc{Herwig~7}'s success in predicting the high-energy precision measurement data from the LHC. This would be a milestone development that can bring forth the next step in the evolution of high-energy Monte-Carlo event generators and can pave the way for further parton-shower-related developments, e.g. BSM parton showers and Higgs boson self-interactions that will become relevant with the upcoming and inevitable push in the probe energies of the existing and future particle colliders. It would also be highly interesting to introduce the same level of enhancement in \textsc{Herwig}'s dipole shower scheme to study the algorithmic behaviors of these methods. Additionally, matching of EW parton shower in the low and high energy scales, between broken and unbroken SU(3)$\times$SU(2)$\times$U(1) gauge theories has to be understood. 

\section{Conclusions} 
\label{sec:conc}

Parton shower event generators have played a central role in the planning and interpretation of experiments since the 1980s. They have, however, moved beyond being simply the QCD inspired models that they were at their inception. Instead, they have evolved into theoretical tools derived from the Lagrangian of the standard model, so that predictions made using these tools can help to test the standard model and potentially reveal new physics beyond the standard model. The algorithms used involve approximations. The present challenge, which is the subject of current theoretical research, is to improve the approximations and thus improve the precision of the resulting predictions. 

\section*{Acknowledgments}

The work of N.~Darvishi is supported by the National Natural Science Foundation of China (NSFC) under grants No. 12022514, No. 11875003 and No. 12047503, and CAS Project for Young Scientists in Basic Research YSBR-006, by the Development Program of China under Grant No. 2020YFC2201501 (2021/12/28) and by the CAS President’s International Fellowship Initiative (PIFI) grant.
M.R.~Masouminia is supported by the UK Science and Technology Facilities Council (grant numbers ST/P001246/1).
The work of D.~Soper was supported by the United States Department of Energy under grant DE-SC0011640.


\end{document}